# Channel Transition Invariant Fast Broadcasting Scheme

Mohammad Saidur Rahman
Department of Computer Science
American International University-Bangladesh
Dhaka, Bangladesh

Ashfaqur Rahman
Department of Computer Science
American International University-Bangladesh
Dhaka, Bangladesh

*Abstract* —Fast broadcasting (FB) is a popular near video-on-demand system where a video is divided into equal size segments those are repeatedly transmitted over a number of channels following a pattern. For user satisfaction, it is required to reduce the initial user waiting time and client side buffer requirement at streaming. Use of additional channels can achieve the objective. However, some augmentation is required to the basic FB scheme as it lacks any mechanism to realize a well-defined relationship among the segment sizes at channel transition. Lack of correspondence between the segments causes intermediate waiting for the clients while watching videos. Use of additional channel requires additional bandwidth. In this paper, we propose a modified FB scheme that achieves zero initial clients waiting time and provides a mechanism to control client side buffer requirement at streaming without requiring additional channels. We present several results to demonstrate the effectiveness of the proposed FB scheme over the existing ones.

**Keywords**-*VoD; FB scheme; channel transition.*

## I. INTRODUCTION

Video-On-Demand (VOD) is a technology [1]-[3] that provides a mechanism to watch a video of choice at will, free of the choice of other viewers. VOD contains numerous applications for instance movie-on-demand, news-on-demand, distance learning, various interactive training programs etc. The appearance of the internet as a persistent communication medium, and a full-grown digital video technology has made VOD a reality. VOD systems are primarily classified into two types: true VOD and near VOD. True VOD systems can provide instantaneous service by allocation of a dedicated channel to each user, whereas near VOD systems perform channel sharing by introducing some service latency. True VOD is achievable with a small number of viewers where a dedicated channel can be provided to everyone. While considering a large population, it becomes necessary to consider near VOD systems. The principal goal of near VOD system is to minimize the average user waiting time.

Fast broadcasting [4]-[9] is a near VOD broadcasting scheme that divides a video into equal size segments whose duration is equal to the maximum user waiting time. These segments are allocated to a small number of channels of equal bandwidth and successive segments are broadcasted in proper decreasing frequencies using time-division multiplexing. While watching the video, the client needs to buffer some future portion of the video as displaying others. The user waiting time and also the client side buffer in FB depends on the number of channels. These two parameters may need to be tuned based on the demand of a video that may change dynamically over a period of time. This requires an increase in the number of channels in FB scheme.

This channel transition however causes lack of correspondence between the segment sizes before and after transition. Some modified FB schemes [1][4] pre-buffer a segment of the original video to create a correspondence between segment sizes at transition. The modified scheme achieves the objective using additional channels and also wasting bandwidth (to be explained in Section II).

In this paper we propose a modified FB scheme that preloads a segment of the video at the client side leading to segment correspondence at channel transition. Unlike the existing approaches, our proposed scheme requires constant number of channels to achieve the abovementioned objective and does not waste any bandwidth. Preloading also leads to zero user waiting time. Analytical results at Section IV reveal the superiority of our proposed technique over the existing ones.

## II. RELATED WORK

In this section we explain the FB scheme. Some modified FB schemes [1][4] deal with the channel transition problem and are explained in this section as well.

## A. Fast Broadcasting Scheme

In FB scheme [4]-[5] $k$ channels namely $C_0, C_1, \ldots, C_{k-1}$ each of bandwidth $b$ work together to broadcast a video $V$ with the following arrangement:

- With the availability of $k$ channels, V is divided into N segments (Figure 1), $S_1^k, S_2^k, \ldots, S_N^k$ where $N = 2^k - 1$ and $S_1^k \bullet S_2^k \bullet \cdots \bullet S_N^k = V$ (•is the concatenation operator). Assuming that the length of the video is D, the length of each segment $\Delta_{FB}^k$ becomes

$$\Delta_{FB}^k = D/N = D/(2^k - 1) \qquad (1)$$

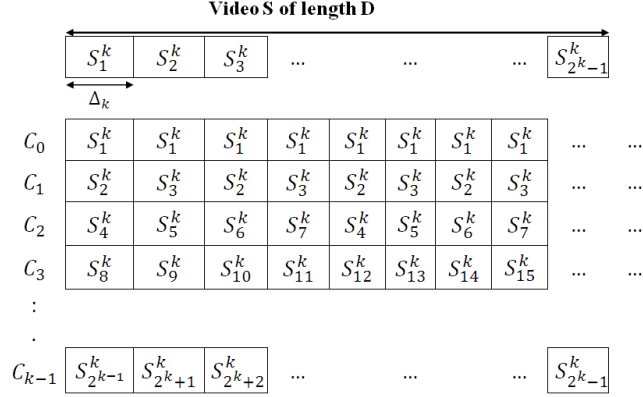

Fig. 1. Fast Broadcasting (FB) Scheme with k channels. The video is segmented into $N = 2^k - 1$ segments $S_1^k \bullet S_2^k \bullet \cdots \bullet S_{2^k-1}^k$ of size $\Delta_{FB}^k = D/(2^k - 1)$. Segments $S_{2^i}^k, S_{2^i+1}^k, \ldots, S_{2^{i+1}-1}^k$ are periodically broadcasted on channel Ci where $i \in \{0, 1, \ldots, k-1\}$.

- With the total number of channels being k, the segments $S_{2^i}^k, S_{2^i+1}^k, \ldots, S_{2^{i+1}-1}^k$ are transmitted periodically on channel $C_i$ (Figure 1) where $i \in \{0, 1, \ldots, k-1\}$.

The initial user waiting time in the FB scheme is proportional to segment size $\Delta_{FB}^k$. The client buffer requirement at the client end can be computed from the number of segments to be stored on the client side while displaying the video. In FB scheme the number of segments downloaded at the client side at $n$–th time instant equals

$$k - \lceil \log_2 n \rceil \qquad (2)$$

where $k$ is the number of channels in use. The total number of downloaded segments up to the $n$–th time instant equals

$$k + \sum_{i=2}^{n} (k - \lceil \log_2 i \rceil) \qquad (3)$$

By that time a total of ($n$–1) segments are displayed. Thus the actual buffer requirement of FB scheme at $n$–th time instant equals

$$buff_{FB} = \left[ k + \sum_{i=2}^{n} (k - \lceil \log_2 i \rceil) - (n-1) \right] \times \Delta_{FB}^k \qquad (4)$$

With $k$ channels the buffer requirement of the FB scheme reaches its maximum at $n = 2^{k-2}$ and remains at the maximum until $n = 2^{k-1}$. The maximum storage requirement of FB scheme thus equals

$$buff_{FB}^{max} = \left[ k + \sum_{i=2}^{2^{k-2}} (k - \lceil \log_2 i \rceil) - (2^{k-2} - 1) \right] \times \Delta_{FB}^k \qquad (5)$$

The problem with the FB scheme is that it only considers the data scheduling of a video given a fixed number of channels. However, it may be required to reduce the initial user waiting time or client side buffer requirement dynamically. The initial client waiting time reduces with incrementing $k$ as observed from (1). As $\Delta_{FB}^k = D/(2^k - 1)$ in (5), the denominator is a polynomial function of *k* whereas the nominator is a linear function of *k*. Thus incrementing *k* reduces client buffer requirement as well in FB scheme.

Changing the number of channels creates a correspondence problem between segment sizes before and after transition in FB scheme. Consider, for instance, the allocation of two and three channels for a video *V* (Figure 2). The length of each segment will be *D*/3 and *D*/7 respectively. Since the denominators are relatively prime, there is no clear correspondence relationship between the segments obtained from these two channel assignments. As a result, if channel transition occurs users need to go back to a previous time slot (when channel transition occurs from *k*=2 to *k*=3). Hence, user needs to re-play some video. User may also need to go to a time slot in front (when channel transition occurs from *k*=3 to *k*=2) hence need download some video as it is not yet available causing *intermediate client waiting*.

*B. Fast Broadcasting Scheme With Data Preloading*

The authors in [1][4] proposes a Data Preloading based Fast Broadcasting (DPFB) scheme that can reduce the initial user waiting time and client side buffer requirement dynamically by addressing the correspondence problem at channel transition. In DPFB, it is assumed that the video *V* is served with at least β channels. Considering a partition of the video *V* into $2^\beta$ equal size non-overlapping segments, it is assumed that the last segment $S^\beta_{2^\beta}$ called $V_{prebuffer}$ is preloaded in a buffer at client side (Figure 3). Assuming that *k* channels $C_0, C_1, ..., C_{k-1}$ are allocated to *V* ($k \geq \beta$) and $S^\beta_{2^\beta}$ is preloaded into client side the FB scheme is modified as follows:

- V is divided into $N = 2^k$ segments $S^k_1, S^k_2, ..., S^k_N$ such that $\sum_{i=1}^{N} S^k_i = D$ where D is the length of the video. The length of each segment is therefore $\Delta^k_{DPFB} = D/2^k$.

- With the number of channels being k, the segments $S^k_{2^i}, S^k_{2^i+1}, ..., S^k_{2^{i+1}-1}$ are transmitted periodically on channel Ci where $i \in \{0,1,...,k-1\}$.

The last segment is never broadcast on any of the channels as is always contains preloaded data at the client side. In DPFB the relationship between segment sizes at channel transition can be defined as

$$\Delta^k_{DPFB} = \frac{D}{2^k} = \frac{D}{2 \times 2^{k-1}} = \frac{1}{2}\Delta^{k-1}_{DPFB}.\tag{6}$$

Thus DPFB resolves the correspondence problem. Following the analysis made for FB scheme, the maximum buffer requirement of DPFB equals

$$buff^{max}_{DPFB} = \Delta^\beta_{DPFB} + \left[k + \sum_{i=2}^{2^{k-2}}(k - \lceil \log_2 i \rceil) - (2^{k-2} - 1)\right] \times \Delta^k_{DPFB}\tag{7}$$

for *k* channels where $\Delta^\beta_{DPFB}$ is the size of the prebuffer.

Fig. 2. Comparison of segments sizes at different number of channels in FB scheme. The figure presents segment sizes for number of channels k=1, 2, and 3.

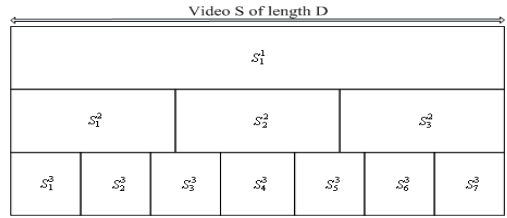

Fig. 3. Relationship among segments at different number of channels in the DPFB scheme with minimum channel requirement β=2.

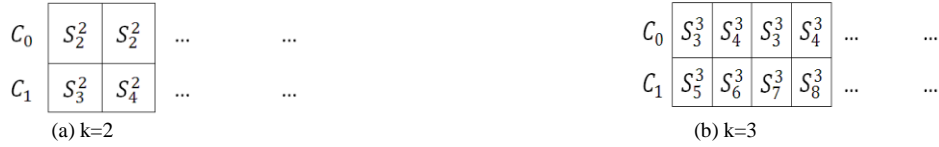

Fig. 4. Relationship among segments at different values of buffer control parameter k in the proposed CTFB scheme. The shaded region refers to already preloaded data. It is assumed that minimum channel requirement γ=2.

|  |  |  |  |  |  |  |  |
|---|---|---|---|---|---|---|---|
| $C_0$ | $S_2^2$ | $S_2^2$ | ... | ... | | | |
| $C_1$ | $S_3^2$ | $S_4^2$ | ... | ... | | | |

(a) k=2

|  |  |  |  |  |  |  |
|---|---|---|---|---|---|---|
| $C_0$ | $S_3^3$ | $S_4^3$ | $S_3^3$ | $S_4^3$ | ... | ... |
| $C_1$ | $S_5^3$ | $S_6^3$ | $S_7^3$ | $S_8^3$ | ... | ... |

(b) k=3

Fig. 5. The transmission of segments in the CTFB scheme at different values of k where # of channels γ=2. Note that at different values of k the pre-buffered segments (Figure-4) are not broadcasted thus requiring constant number of channels.

One disadvantage with DPFB is that some data segments are transmitted even if they are already preloaded (e.g. $S_3^7$ with k=3 in Figure 5 has to be transmitted even though it is already preloaded at client side). This leads to wastage of bandwidth.

## III. PROPOSED VOD SCHEME

The objective of the proposed method is to setup an arrangement so that the objective of reduced initial user waiting time and dynamic control of client side buffer requirement is possible without requiring additional channels. Our proposed Channel Transition invariant Fast Broadcasting (CTFB) scheme is based on the following foundations:

- It is assumed that a video V is always served with γ channels.

- Considering a partition of the video V into $2^γ$ equal size non-overlapping segments, it is assumed that the first segment $V_{PB}$ (Figure 4) is preloaded in a buffer at the client side.

- There is a client buffer control parameter k. With k the video V is divided into $2^k$ segments.

Considering that $V_{PB}$ is preloaded into client side buffer, γ channels $C_0, C_1, ..., C_{γ-1}$ allocated to V, and client buffer control parameter k the proposed CTFB scheme works on the following arrangement:

- V is divided into $N = 2^k$ segments $S_1^k, S_2^k, ..., S_N^k$ such that $\sum_{i=1}^{N} S_i^k = D$ where D is the length of the video. The length of each segment is therefore $\Delta_{CTFB}^k = D/2^k$.

- The segments $S_{2^{k-γ+i}+1}^k, S_{2^{k-γ+i}+2}^k, ..., S_{2^{k-γ+i+1}}^k$ are transmitted periodically (Figure 5) on channel $C_i$ where i ∈ {0,1,...,k−1}.

Note that the segments $S_1^k, S_2^k, ..., S_{2^{k-γ+i}}^k$ are never broadcasted in our proposed scheme as they are already preloaded at the client side buffer (Figure 4). With CHFB scheme we achieve the followings:

- The preloaded segment VPB is equal to $S_1^k \bullet S_2^k \bullet \cdots \bullet S_{2^{k-γ+i}}^k$ for any k. As the number of segments to be transmitted starts from $S_{2^{k-γ+i}+1}^k$, CTFB thus do not broadcast already preloaded segments (the shaded region in Figure 4). Under identical scenario preloaded segments are transmitted in DPFB (Figure 3) scheme leading to wastage of bandwidth.

- As the first segment VPB is preloaded at the client buffer, the user can start watching the video at any time. The remaining segments are always available on the different channels transmitted from the server. This leads to zero initial user waiting time in the proposed CTFB scheme. In DPFB scheme the initial user waiting time is proportional to segment size $\Delta_{DPFB}^k$.

- There exists a definite relationship among the segments at change of k. As

$$\Delta_{CTFB}^k = \frac{D}{2^k} = \frac{D}{2 \times 2^{k-1}} = \frac{1}{2} \Delta_{CTFB}^{k-1}, \qquad (8)$$

There is a precise correspondence between the segment sizes at the change of client buffer control parameter $k$. The maximum buffer requirement at client side in CTFB at $n$–th time instant equals

$$buff_{CTFB}^{max} = \Delta_{CTFB}^{\gamma} + \left[ k + \sum_{i=2}^{2^{k-2}} (k - \lceil \log_2 i \rceil) - (2^{k-2} - 1) \right] \times \Delta_{CTFB}^{k}. \quad (9)$$

It can be observed that changing $k$ will change the segment size that thus the buffer requirement. $k$ is thus called the client buffer control parameter.

- As the segment size is controlled by client buffer control parameter $k$ number of channel transition is required in CTFB leading to reduced use of bandwidth. The situation is clarified in Figure 5.

IV. RESULTS AND DISCUSSION

In this section we present some results showing the performance of CTFB and at the same comparing with FB, DPFB and CTFB scheme. For analysis we have considered a video of size 10 MB and client playback rate of 10 kbps. The results are reported from Figure 6-Figure 9. The features of CTFB and relevant discussion are presented next:

*A. Zero user waiting time*

In the CTFB scheme there is no user waiting for the client. When a user starts watching a video the user is served instantly with the preloaded segments from the client end buffer. The remaining segments are downloaded from the channels. As the user can start instantly it leads to zero user waiting time. Figure 6 shows the comparison of the waiting time for user of FB, DPFB and CTFB scheme. It can be observed that the user waiting time in FB and DPFB decreases with increased number of channels whereas in CTFB it is always zero.

*B. Channel requirement*

In CTFB the segment size (and thus client side buffer requirement) is controlled by a parameter $k$. The number of channel is always $\gamma$ and the server is free of the burden for managing additional bandwidth dynamically. In FB and DPFB the segment size (and thus user waiting time and buffer) is reduced by increasing the number of channels. This requires server overhead for managing extra bandwidth. A comparison of the required number of channels at different segment sizes is present in Figure 7. Note that the number of channels in CTFB is constant whereas is in FB and DPFB channel requirement goes down with increased segment size.

*C. Reduced bandwidth requirement*

In DPFB scheme, the segments $S_{2^i}^k, S_{2^i+1}^k, \ldots, S_{2^{i+1}-1}^k$ are periodically broadcasted through $C_i$ whereas in the CTFB scheme the segments $S_{2^{k-\gamma+i}+1}^k, S_{2^{k-\gamma+i}+2}^k, \ldots, S_{2^{k-\gamma+i+1}}^k$ are periodically transmitted through channel $C_i$, where $i \in \{0, 1, \ldots, k-1\}$. Fewer segments need to be transmitted in CTFB. With increased number of channels this difference is even bigger. This is evidenced from the comparison of the number of segments to be transmitted for FB, DPFB and CTFB scheme in Figure 7.

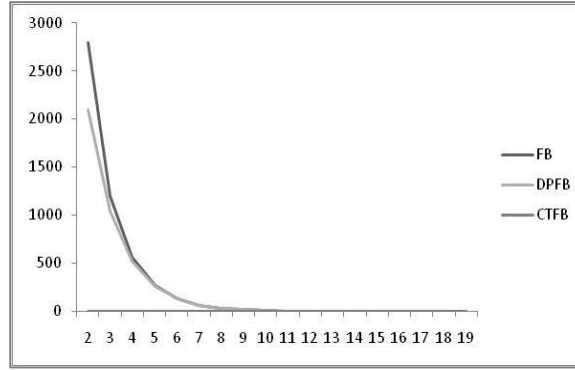

Fig. 2. Comparison of user waiting time (in y-axis) for FB, DPFB and CTFB scheme for different number of channels $k$ (in x-axis).

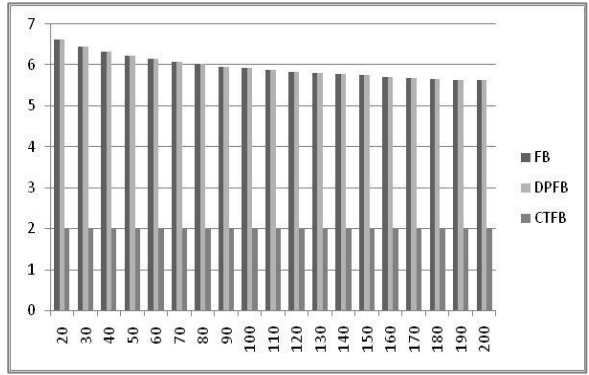

Fig. 3. Comparison of the channel requirement (in y-axis) for FB, DPFB and CTFB scheme at different values of segment size (in x-axis).

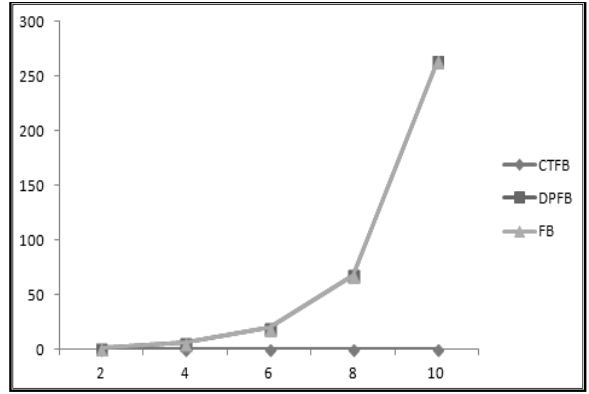

Fig. 4. Comparison of the number of redundant segments to be transmitted (in y-axis) for the CTFB, FB and DPFB scheme for different number of channels (in x-axis). For CTFB it is assumed that $\gamma=2$ and for FB and DPFB $\beta=2$.

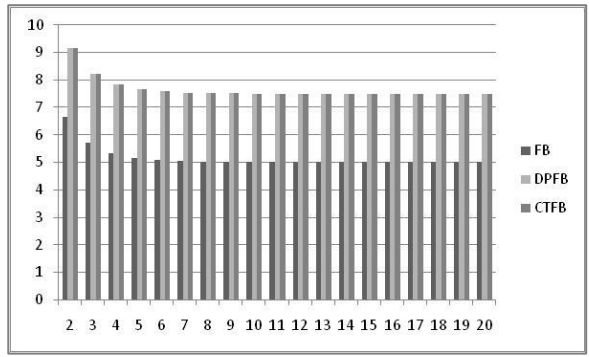

Fig. 5. Comparison of buffer requirement in MB (in y-axis) for the FB, DPFB and CTFB scheme at different values of k (in x-axis). Here $\gamma=\beta=2$.

*D. Zero bandwidth wastage*

In CTFB scheme the preloaded segments are not transmitted leading to no wastage of bandwidth. In DPFB scheme the segments $S^k_{(2^\beta-1)\times 2^{k-\beta}}, \ldots, S^k_{2^k-1}$ are transmitted even though they are preloaded at the client side buffer (the shaded region in Figure 4). The number of redundant channels to be transmitted increases exponentially. This is portrayed in Figure 8. Note that in FB scheme the redundant segments are transmitted.

*E. Same buffer required as DPFB*

The client side buffer requirement in CTFB depends on $k$ and $\gamma$ as evidenced from (9). In DPFB as well the size of the buffer depends on the value of $\beta$ and $k$. For the sake of fairness, if the same values are assumed for both of these parameters, the buffer requirement for FB, DPFB, and CTFB is presented in Figure 9. Note that the buffer requirement for DPFB and CTFB scheme is same but is higher than FB scheme. As FB scheme does not preload any data it requires fewer buffers. Also note that CTFB can control the buffer size at client end with $k$. The higher the value of $k$ the lower buffer required at the client side.

## V. Conclusion

In this paper we have presented a modified Fast VOD scheme that is capable of serving at constant number of channels while achieving the objective of reduced bandwidth and providing a parameter to control client end buffer size. As it does not transmit preloaded video segments CTFB does not waste bandwidth. As users can start from the buffer instantly the user waiting time becomes zero. Like DPFB scheme CTFB also imposes the constraint of preloading some part of the video beforehand.

[1] S. A. Azad, "New strategies for efficient video-on-demand systems," PhD thesis, GSIT, Monash University, Australia 2007.

[2] H. –F. Yu, H. –C. Yang, C. –Y. Chien, and Y. –T. Wang, "A Limited-Client-Capability Broadcasting Scheme for VoD Applications," Seventh International Conference on Networking, pp. 192–196, April 2008.

[3] J. Chen and J. –H. Sun, "Low buffering and waiting-time video-on-demand broadcasting scheme for WiMAX systems," ACM International Conference on Mobile multimedia communications, article no. 40, 2007.

[4] L.-S. Jhun and L.-M. Tseng, "Fast data broadcasting and receiving for popular video service," IEEE Transactions on Broadcasting, vol. 44, no. 1, pp. 100-105, March 1998 .

[5] J.-F. Paris, "A simple low-bandwidth broadcasting protocol for video-on-demand," in Proc. Of International Conference on Computer Communication and Network, pp. 118-123, 1999.

[6] S. A. Azad and M. M. Murshed, "Seamless channel transition for popular video broadcasting", Information Technology: Coding and Computing, 2005. ITCC 2005. International Conference on Volume 2, pp. 283 – 288, April 2005.

[7] Y.-C. Tseng, M.-H. Yang, C.-M. Hsieh, W.-H. Liao, and J.-P. Sheu, "Data broadcasting and seamless channel transition for highly-demanded videos," IEEE Transactions on Communications, vol. 49, no. 5, pp. 863-874, May 2001.

[8] M.-H. Yang, C.-H. Chang, and Y.-C. Tseng, "A borrow-and-return model to reduce client waiting time for broadcasting-based VOD services," IEEE Transactions on Broadcasting, vol. 49, no. 2, pp. 162-169, June 2003.

[9] D. A. Tran, K. A. Hua, and S. Sheu, "A new caching architecture for efficient video-on-demand services on the Internet," in Proc. of the 2003 Symposium on Applications and the Internet (SAINT' 03), pp. 172-181, January 2003.